\documentclass[]{IEEEtran} 
\usepackage[T1]{fontenc}  
\usepackage[italian,french,english]{babel}
\usepackage{amsmath,amsfonts,amssymb,bm,mathrsfs,upgreek} 
\usepackage{graphicx,color}
\usepackage[dvipsnames]{xcolor}
\usepackage{url}

\newcommand{\degrees}{\ensuremath{\mathrm{^\circ}}}   
\newcommand{\unit}[1]{\ensuremath{\mathrm{#1}}}       
\graphicspath{{./Figures/}}

\IEEEoverridecommandlockouts 
\DeclareRobustCommand*{\IEEEauthorrefmark}[1]{%
  \raisebox{0pt}[0pt][0pt]{\textsuperscript{\footnotesize\ensuremath{#1}}}}
\begin{document}
\bstctlcite{IEEEexample:BSTcontrol}
\selectlanguage{english}


\title{\LARGE Vocabulary, Physical Quantities and Units\\for the Measurement of Amplitude Noise and Phase Noise}
\author{%
  \IEEEauthorblockN{%
    Enrico Rubiola\IEEEauthorrefmark{1,}\IEEEauthorrefmark{2},
    Jacques Millo\IEEEauthorrefmark{1,}\IEEEauthorrefmark{2},
    Nora Meyne\IEEEauthorrefmark{3}, 
    Joseph Achkar\IEEEauthorrefmark{4}, 
    Filippo Levi\IEEEauthorrefmark{5},
    Archita Hati\IEEEauthorrefmark{6}
  } \\
  \IEEEauthorblockA{%
    \IEEEauthorrefmark{1}LNE Laboratoire Temps Fréquence de Besançon (LNE-LTFB), France\\
    \IEEEauthorrefmark{2}Université Marie et Louis Pasteur, SUPMICROTECH, CNRS, Institut FEMTO-ST, Besançon, France\\
    \IEEEauthorrefmark{3}Physikalisch-Technische Bundesanstalt (PTB), Braunschweig, Germany\\
    \IEEEauthorrefmark{4}Chair of the CCTF Working Group on CIPM MRA(BIPM), and LTE, Paris, France\\
    \IEEEauthorrefmark{5}Istituto Nazionale di Ricerca Metrologica (INRiM), Torino, Italy\\
    \IEEEauthorrefmark{6}(6) NIST, Boulder, CO, USA
  } \\[1ex]
  Corresponding author\!\!: Enrico Rubiola, mail enrico.rubiola@femto-st.fr, home page \url{https://rubiola.org}
}

\maketitle
\IEEEpubidadjcol

\begin{abstract}
  The widespread use of non-SI quantities and units, together with improper and misleading terminology, generates confusion in the domain of phase and amplitude noise.  We discuss such physical quantities, units and terms with the purpose of stimulating a discussion between labs, and to agree for the full adoption of the International System of Units (SI) and clear, unambiguous terminology.
\end{abstract}


\pagenumbering{gobble}

\section{Introduction}
The terms `phase noise' and `amplitude noise' refer to the random processes $\varphi(t)$ and $\alpha(t)$ affecting a clean periodic signal of amplitude $V_0$ and frequency $f$ like
\begin{equation}
  v(t)=V_0\bigl[1+\alpha(t)\bigr]\cos\bigl[2\pi f t+\varphi(t)\bigr].
\end{equation}
Phase noise, PM noise and PN are often used interchangeably, and amplitude noise, AM noise and AN as well.  Phase noise measurements are generally represented in the spectral domain but, for historical reasons having roots in the 1964 IEEE-NASA Symposium \cite{Chi:1965short}, the quantity \(\mathscr{L}(f)\) is preferred to the power spectral density (PSD) \(S_{\varphi}(f)\) of the phase \(\varphi(t)\).
The quantity \(\mathscr{L}(f)\), given in dBc/Hz, is adopted by all manufacturers of phase noise analyzers and RF/microwave oscillators, with no known exception, and is widely used in the literature.  Oddly, \(\mathscr{L}(f)\) and dBc/Hz are also used for AM noise, albeit \(\mathscr{L}(f)\) is sometimes replaced with \(\mathscr{M}(f)\).  The reader is encouraged to refer to Refs.\ \cite{Donley:2022shortest,Rubiola-2023-Companion} for a review.

There is a lot of confusion in the domain of PM and AM noise, related to the wide use of improper and misleading vocabulary, and to the the use of the non-SI quantity \(\mathscr{L}(f)\) and unit dBc/Hz.  Under the coordination of BIPM, phase noise goes with `Electricity and Magnetism,' but most of the scientific advances relate to `Time and Frequency.'

After the major revision of the SI in 2019, and with the re-definition of the SI second in progress \cite{Dimarcq-2024}, this work proposes to start a revision of quantities, units, and terminology to be consistent with the SI.  We encourage other metrological institutes to adhere to this initiative.

\section{Physical Quantities and Units Consistent with the SI}
Inside all phase noise analyzers, the measurement starts with the measurement of the random phase \(\varphi(t)\), and the single-sided PSD is evaluated with the Welch algorithm as
\begin{equation}
  S_\varphi(f) = \frac{2}{T} \Bigl<\bigl|\Phi_T\bigr|^2\Bigr>_m, \qquad f>0
\end{equation}
where \(\Phi_T(f)\) is the discrete Fourier transform of \(\varphi(t)\) truncated over a suitable measurement time \(T\), the brackets \(\langle \ldots \rangle_m\) denote the average on a suitable number \(m\) of acquisitions, and the factor 2 accounts for energy conservation after deleting the negative frequencies.  The angle has physical dimension of 1 and the associated SI unit is the radian, symbol rad.  Thus, \(S_\varphi(f)\) has the physical dimension of time, and the associated SI unit is \unit{rad^2\,s}, or equivalently \unit{rad^2/Hz}.  The decibel, symbol dB, is a non-SI unit allowed for use with SI units, and particularly useful to us because \(\mathscr{L}(f)\) often spans over multiple decades.  In conclusion, the quantity \(S_\varphi(f)\) and the associated unit \unit{rad^2/Hz}, or \unit{dB\,rad^2/Hz}, are the obvious choice because they match the measurement process, and they are consistent with the SI\@.  The fractional amplitude \(\alpha(t)\) has the dimension of 1, and no associate measurement unit.  Thus, \(S_\alpha(f)\) and the associated unit \unit{dB\,1/Hz} (or \unit{dB/Hz}) seems a natural choice, but \unit{dB\,(V/V)^2/Hz} may also be considered for use in electrical engineering.

\section{The Weird Quantity \(\mathscr{L}(f)\)}
\begin{figure}[t]   
  {\centering\includegraphics[width=0.8\columnwidth]{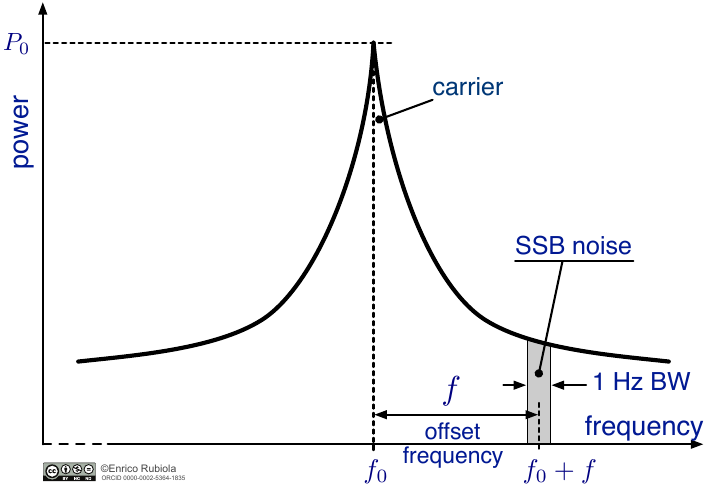}\par}
  \caption{Widely used, and misleading interpretation of \(\mathscr{L}(f)\) in terms of 1-Hz noise to carrier ratio.  \includegraphics[height=1em]{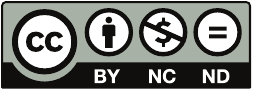}~Enrico~Rubiola.}  
  \label{fig:Old-L}
\end{figure}
The quantity \(\mathscr{L}(f)\) is improperly called `SSB noise' (Fig.~\ref{fig:Old-L}), and the definition
\begin{equation}
  \mathscr{L}(f) = \frac{\text{\small noise power in 1 Hz bandwidth at the offset \(f\)}}{\text{\small carrier power}}
  \label{eqn:Old-L}
\end{equation}
is still found in numerous documents. \(\mathscr{L}(f)\) is given in dBc/Hz using \(10\log_{10}\) dB, where `c' stands for `referred to the carrier power.'

A first problem is that the SSB noise does not say if the noise is AM or PM, or any combination of.  In fact, modulation goes with specific relations between upper and lower sideband (LSB and USB), but a single sideband does not say anything about such relations.  The term `offset frequency' reinforces the wrong belief that a single sideband is sufficient to define the phase noise.

A second problem is that the generally agreed definition
\begin{equation}
  \mathscr{L}(f) = \tfrac{1}{2}\,S_\varphi(f)
\end{equation}
means that \(\mathscr{L}(f)\) and \(S_\varphi(f)\) are the same thing, but differ in the measurement units (factor 1/2), which is inconsistent with Fig.~\ref{fig:Old-L}.  Starting from the usual expression in decibels
\begin{gather}
  \mathrm{unit}\bigl\{10 \log_{10}\big[\mathscr{L}(f)\big]\bigr\} = \unit{dBc/Hz}\\
  \mathrm{unit}\bigl\{10 \log_{10}\big[S_\varphi(f)\big]\bigr\} = \unit{dB\,rad^2Hz}
\end{gather}
and stripping the `dB' and `Hz' units, we realize that \(\mathscr{L}(f)\) is a square angle
\begin{equation}
  1\unit{c} = 2\unit{rad^2} 
\end{equation}
and consequently, the unit of angle used in \(\mathscr{L}(f)\) is
\begin{gather}
  \sqrt{\unit{c}} = \sqrt{2}\unit{rad} \simeq 81\degrees~. 
\end{gather}

\begin{figure}[t]   
  {\centering\includegraphics[width=0.98\columnwidth]{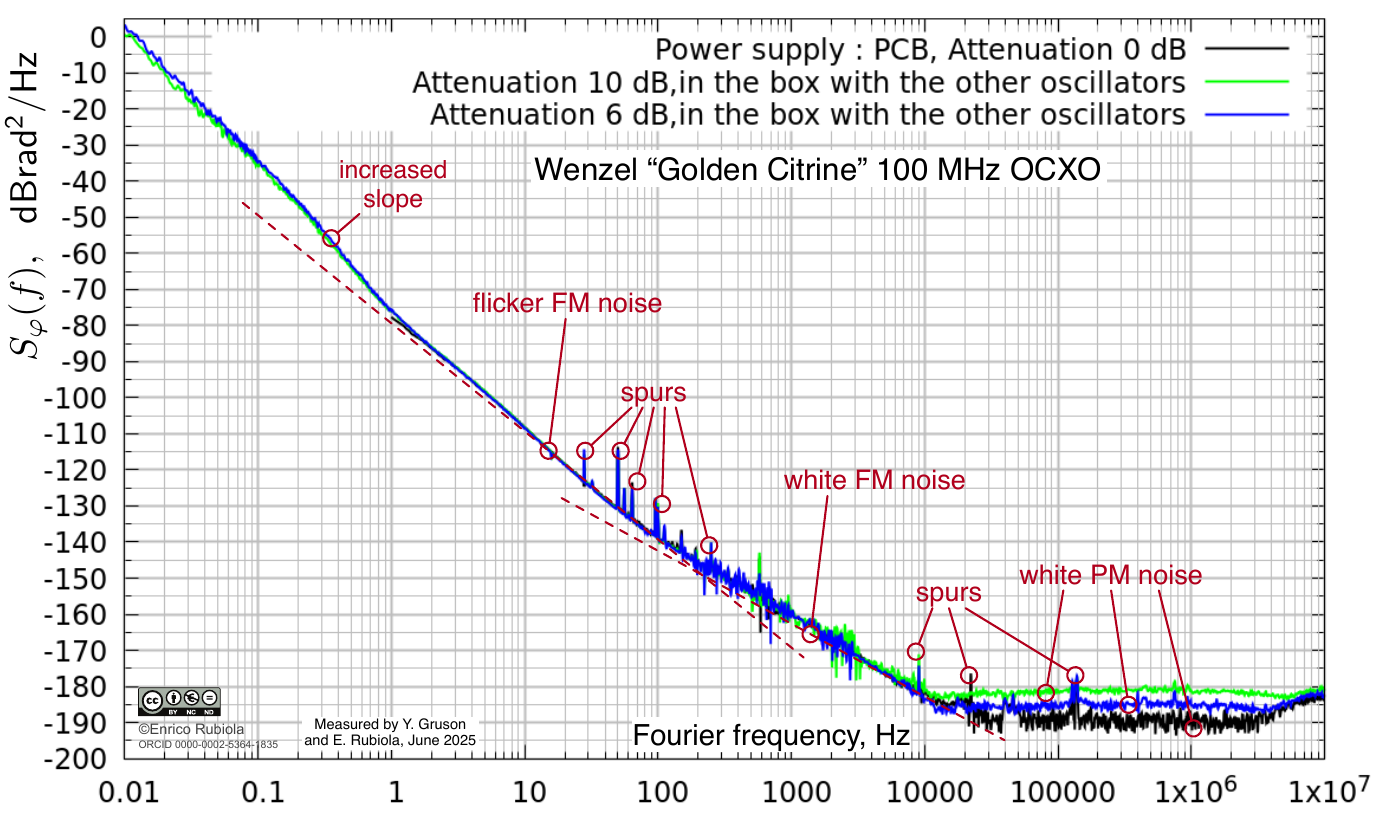}\par}
  \caption{Example of phase noise measurement of an oscillator (100 MHz OCXO) in different experimental conditions. \includegraphics[height=1em]{by-nc-nd}~Enrico~Rubiola.}  
  \label{fig:Sphi-Citrine}
\end{figure}

A third problem is that the SSB noise can only be used to describe small amounts of phase noise.  In fact, angular modulations are ruled by the energy conservation, thus the sideband power comes at expenses of the carrier power.  When the sideband power is not negligible, \eqref{eqn:Old-L} breaks down. This is the case of the noise processes always found in actual oscillators, where \(\varphi(t)\) diverges in the long run (see Fig.~\ref{fig:Sphi-Citrine}).

\section*{Vocabulary and Other Issues}
Besides physical quantities and units, we have observed that some terms need to be clarified or changed, chiefly
\begin{itemize}
  \item \emph{Parametric noise} may be a too exotic term, not easily identified as a random modulation
  \item \emph{Additive noise} and \emph{random noise} often raise confusion when it is discussed together with the parametric noise
  \item \emph{Added noise} (of a two-port device). This is the noise of the device, and there is no reason for saying ``added''
  \item \emph{Offset frequency} is not suitable to describe a modulation
\end{itemize}
Uncertainty and zero uncertainty are other relevant questions, but they are too wide and will go in a separate work.

\section*{Related Work}
The first international comparison of oscillator phase noise in progress \cite{Rubiola-Pilot}, under the leadership of LNE-LTFB\@.  The questions we raise here will inevitably be addressed during this comparison.  The consistency of phase noise measures with SI and metrological practice is also under discussion at the EURAMET TC-TF \cite{Rubiola.2026a}.  PTB has started the TransMeT project on traceability and comparisons of phase noise measurements \cite{Meyne-TransMeT}.

\section*{Calibration and Measurement Capabilities}
The LNE is the one and only laboratory having phase-noise Calibration and Measurement Capabilities (CMCs) in the BIPM Key Comparison Data Base (KCDB), and no lab has AM noise CMCs yet.  Such CMCs use the quantity \(mathscr{L}\) and the unit dBc/Hz, which are not consistent with the SI (though no objection was raised during the review process) and need to be rewritten from scratch.  Showing this mistake, we hope to catch the attention of other laboratories interested in phase noise and raise a scientific discussion.

\section*{Acknowledgements}
This project is funded by LNE \url{https://www.lne.fr/en}, by the French FIRST-TF network \url{https://first-tf.com/}, by the EIPHI Graduate school (Grant ANR-17-EURE-0002), and by internal funds from Oscillator IMP platform.

\bibliographystyle{IEEEtran} 

\end{document}